\documentclass[onecolumn]{svjour3}
\usepackage{graphicx,subfigure,amssymb,multirow,lscape,pifont,fix-cm}
\input{Qcircuit.tex}
\renewcommand{\Qcircuit}[1][0em]{\xymatrix @*=<#1>}
\usepackage{rotating}
\usepackage{color}
\smartqed
\begin{document}

\thispagestyle{empty}

\clearpage

\setcounter{page}{1}

\title{Synthesis of Quantum Circuits for Linear Nearest Neighbor Architectures}
\author{
 Mehdi Saeedi$^\S$, Robert Wille$^\dag$, Rolf Drechsler$^\flat$ \\ \\
\small $^\S$ Computer Engineering Department, Amirkabir University of Technology, Tehran, Iran\\
\small E-mail: msaeedi@aut.ac.ir \\\\
\small $^\dag$ Institute of Computer Science, University of Bremen, Bremen, Germany\\
\small E-mail: rwille@informatik.uni-bremen.de \\\\
\small $^\flat$ Institute of Computer Science, University of Bremen, Bremen, Germany\\
\small E-mail: drechsler@uni-bremen.de \\
}
\date{}
\authorrunning{M. Saeedi, R. Wille, R. Drechsler}

\maketitle
\newcommand{\MS}[1]{\textcolor{red}{{#1}}}

\begin{abstract}
While a couple of impressive quantum technologies have been proposed, they have several intrinsic limitations which must be considered by circuit designers to produce realizable circuits. Limited interaction distance between gate qubits is one of the most common limitations. In this paper, we suggest extensions of the existing synthesis flow aimed to realize circuits
for quantum architectures with linear nearest neighbor (LNN) interaction. To this end, a template matching optimization, an exact synthesis approach, and two reordering strategies are introduced. The proposed methods
are combined as an integrated synthesis flow. Experiments show that by using the suggested flow, quantum cost can be improved by more than 50\% on average.
\end{abstract}

\section{Introduction}
Since the invention of the integrated circuit in 1958, the number of transistors in such circuits has doubled approximately every two years (also known as Moore's Law). Currently, semiconductor technology has advanced the world towards more powerful systems by decreasing the transistor size. However, further miniaturization is beginning to appear insoluble due to the density of power dissipation and the impossibility to realize patterning features approaching the atomic scale.

The difficult barriers to the ongoing improvements in semiconductor technology have intensified the attraction of alternative computing paradigms such as quantum computing. It has been shown that quantum computing could improve the rate of advance in processing power at least for several applications \cite{Nielsen00}. In principle, there are several problems that cannot be executed on a classical Turing machine as efficiently as on a quantum computer. Quantum computers would provide exponential speedups on several problems including factoring of numbers and simulating the quantum-mechanical behavior of physical systems \cite{Mosca}. However, several obstacles exist in the way of physically implementing scalable quantum computers.

While several impressive physical realizations have been proposed for quantum computers (see \cite{Meter06} for a classification scheme of different quantum computing technologies), all of these technologies have serious intrinsic limitations
\cite{Ross08}.
Among the different technological constraints, limited interaction distance between gate qubits is one of the most common ones. Although arbitrary-distance interaction between qubits is possible in quantum computer technologies with moving qubits (for example in a photon-based system \cite{Photon}), restrictions exist in other quantum technologies. In fact, many physical quantum computer proposals only permit interactions between adjacent (nearest neighbor) qubits \cite{FDH:2004}. For example, trapped ions (e.g.,~\cite{TrappedIon}), liquid nuclear magnetic resonance (NMR) (e.g.,~\cite{NMR}), and the original Kane model \cite{Kane} have been designed based on the interactions between linear nearest neighbor (LNN) qubits. The LNN architecture is often considered as an appropriate approximation to a scalable quantum architecture. If one can show that a circuit can efficiently be realized using an LNN architecture, it can be run  in many other architectures as well \cite{Maslov07}.

The efficient realization of a given quantum algorithm for the LNN architectures is an active research area.
In the recent years, the effect of restricted interactions on several specific quantum algorithms has been studied.
For example, the physical implementation of the quantum Fourier transformation (QFT) \cite{Takahashi:2007}, Shor's factorization algorithm \cite{FDH:2004,Kutin:2007}, quantum addition \cite{Choi08}, and quantum error correction \cite{Fowler} for the LNN architectures have been explored in the past. Besides that, researchers also considered the effects of LNN architectures on the synthesis of general quantum/reversible circuits. In \cite{mottonen06} and \cite{Shende06}, the worst-case synthesis cost of a general unitary matrix under the nearest neighbor restriction has been discussed.
It has been shown that restricting CNOT gates to nearest neighbor interactions increases CNOT count of \cite{Shende06} by at most a factor of 9.
The authors of \cite{Cheung07} showed that translating an arbitrary circuit to the LNN architectures requires a linear increase in the quantum cost with respect to the number of qubits. In \cite{Chakrabarti07,Khan08,Hirata09}, heuristic methods for converting an arbitrary circuit to its equivalent on the LNN architectures have been proposed. However, their performance is limited as discussed later.

In this paper, we suggest extensions of the existing synthesis flow aimed to realize circuits for LNN architectures. We show that with a naive treatment of the LNN restriction, quantum circuits require up to one order of magnitude higher quantum cost in the LNN architectures. In contrast, if this restriction is explicitly considered by the proposed synthesis flow, this increase can be reduced by more than 50\% on average (83\% in the best case). To this end, the following approaches are proposed:

\begin{itemize}
\item An improved template-matching post-synthesis optimization method that reduces the circuit cost for LNN architectures, 
\item an exact synthesis method for small functions realizing circuits with nearest neighbor interaction, and 
\item reordering strategies, which modify the initial qubit locations in order to reduce the distance between non-neighbored qubits.
\end{itemize}

The remainder of this paper is organized as follows. In Section~\ref{sec:prelim}, basic concepts are introduced. Next, we briefly review the naive synthesis flow for LNN architectures in Section~\ref{sec:nnc_synth}. Followed by this, Section~\ref{sec:opt} describes the proposed synthesis and optimization approaches with explicit consideration of the LNN limitation in detail.
How to combine the respective approaches as an integrated flow is sketched in Section~\ref{sec:integration}.
Finally, experimental results are given in Section \ref{sec:exp} and conclusions are drawn in Section~\ref{sec:conclusion}, respectively.
\section {Background} \label {sec:prelim}
\subsection{Reversible Logic}
A function $f:\mathbb{B}^n\rightarrow \mathbb{B}^n$ over variables $X=\{x_1,\dots , x_n\}$ is reversible if it maps each input assignment to a unique output assignment. Such function must have the same number of input and output variables. In this paper, $n$ is particularly used to refer to the number of inputs/outputs.
A circuit realizing a reversible function is a cascade of reversible gates.
Common reversible gates include:
\begin{itemize}
\item A {\em multiple control Toffoli gate} $t_m$ has the form $t_m(C, t)$, where $C = \{x_{i_1} , \dots , x_{i_m}\}$ $\subset X$ is the set of \emph{control lines} and $t = \{x_j\}$ with $C \cap t = \emptyset$ is the \emph{target line}. The value of the target line is inverted iff all control lines are assigned to~1. For $m$=0 and $m$=1, the gates are called \textit{NOT} gate and \textit{CNOT} gate, respectively. For $m$=2, the gate is called \textit{C$^2$NOT} gate or \textit{Toffoli} gate.
\item A {\em multiple control Fredkin gate} $f_m$ has two target lines and $m$ control lines. The gate interchanges the values of the target lines iff the conjunction of all $m$ control lines evaluates to 1. For $m$=0, the gate is called \textit{SWAP} gate.
\item A {\em Peres gate} $P$ has one control line $x_i$ as well as two target lines $x_{j_1}$ and $x_{j_2}$. It represents a $t_2(\{x_{i},x_{j_1}\}, x_{j_2})$ and a $t_1(\{x_{i}\}, x_{j_1})$ in a cascade.
\end{itemize}
Reversible logic has applications in various fields including quantum computation.

\newpage

\subsection{Quantum Logic}
A quantum bit, qubit in short, 
can be realized by a physical system such as a photon. Each qubit has two basic states $|0\rangle$ as well as $|1\rangle$ and can get any linear combination of its basic states (called superposition, as shown in (\ref{Eq:dirac}) where $\alpha$ and $\beta$ are complex numbers).
\begin{equation}\label{Eq:dirac}
|\psi\rangle = \alpha|0\rangle+\beta|1\rangle
\end{equation}

An $n$-qubit quantum gate is a device which performs a specific $2^n\times2^n$ unitary operation on selected $n$ qubits in a specific period of time. A matrix $U$ is unitary if $UU^\dag=I$ where $U^\dag$ is the conjugate transpose of $U$ and $I$ is the identity matrix. Previously, various quantum gates with different functionalities have been introduced. For examples, Hadamard (H), Controlled-V, and Controlled-V$^+$ gates are defined by the following unitary matrices:
\begin{equation}
\begin{array}{l}
  H  = \frac{1}{{\sqrt 2 }}\left[ {\begin{array}{*{20}c}
   1 & 1  \\
   1 & { - 1}  \\
\end{array}} \right]
 \end{array}
\label{gate:H}
\end{equation}

\begin{equation}
\begin{array}{l}
  V  = \frac{1+i}{2}\left[ {\begin{array}{*{20}c}
   1 & -i  \\
   -i & 1  \\
\end{array}} \right]
 \end{array}
\label{gate:V}
\end{equation}

\begin{equation}
\begin{array}{l}
  V^+  = \frac{1-i}{2}\left[ {\begin{array}{*{20}c}
   1 & i  \\
   i & 1  \\
\end{array}} \right]
 \end{array}
\label{gate:V+}
\end{equation}
\subsection{Synthesis Cost} \label {sec:sync_cost}
Each Toffoli, Fredkin, and Peres gate can be \emph{decomposed} into a quantum circuit composed of a sequence of \emph{elementary quantum gates} \cite{Nielsen00}. Each elementary gate performs a single physical operation in a certain quantum computing technology. The number of elementary gates required to realize a given reversible gate is called \emph{quantum cost}. It has been shown that NOT, CNOT, Controlled-V, and Controlled-V$^+$ gates can efficiently be realized in quantum computer technologies \cite{Lee06}. These gates are usually considered as elementary gates for reversible Boolean functions \cite{Barenco95}. Thus, we stay with this definition in the following sections. However, in other technologies not only this restricted set, but all one-qubit gates and all two-qubit gates, respectively, are considered as elementary gates~\cite{Nielsen00}. This is separately considered in the experimental evaluation of the proposed approach in Section~\ref{sec:exp}.

\begin{figure}
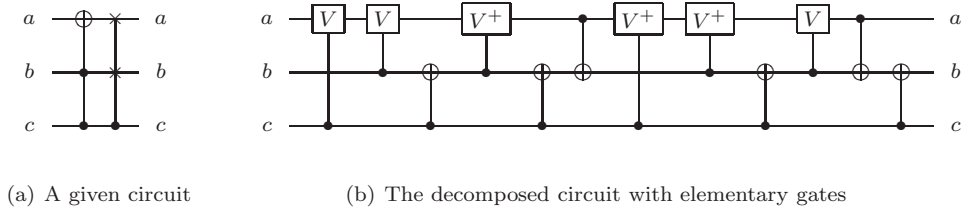

    \centering
        \subfigure[A given circuit \label{fig:example_rev}]{
            \input{qcircuits/example_rev.tex}
        }
    \subfigure[The decomposed circuit with elementary gates\label{fig:example_decomp}]{
        \input{qcircuits/toffoli_2_decom.tex}
    }
    \caption{A reversible circuit and its decomposed circuit}
\end{figure}

Fig.~\ref{fig:example_rev} shows a Toffoli gate and a Fredkin gate in a cascade. The resulting (decomposed) quantum circuit is depicted in Fig.~\ref{fig:example_decomp}. Here, the control lines are denoted by \ding{108} while the target lines are denoted by $\oplus$, $\times$, a V box, or a V$^+$ box, respectively.
As can be seen, a $t_2$ gate is decomposed into 5 elemantary gates, while a $f_m$ gate is decomposed into 7 elemenary gates, respectively. For larger gates, the respective decomposition depends on the number $n-m$ of unused circuit lines:
For $n \geq 5$ and $m \in \{3,4, \cdots \left\lfloor n/2 \right\rceil\}$, a $t_m$ gate can be decomposed into a linear-size circuit which contains $12m-22$ elementary gates. In addition, for $n \geq 7$, a $t_{n-2}$ gate can be decomposed into $24n-88$ elementary gates with no auxiliary bits \cite{MaslovTCAD08}. Finally, a t$_{n-1}$ gate can be decomposed into $2^n-3$ elementary gates if no unused circuit line is available \cite{Barenco95}. The cost of a $f_m$ gate ($1\leq m \leq n-2$) is the cost of a $t_{m+1}$ gate plus two \cite{Nielsen00}. Obviously, always the most efficient decomposition is applied.

\section {The Naive Synthesis Flow for the LNN Architectures} \label {sec:nnc_synth}

Reversible circuits can be synthesized using multiple control Toffoli gates first that are afterwards mapped to elementary quantum gates. On the other hand, elementary gates can be directly applied during the synthesis process. While for the latter case, only small circuits have been determined so far (e.g.,~see \cite{Hung06,GrosseMVL08}), approaches for Toffoli network synthesis can handle larger functions and circuits (e.g.,~see~\cite {MDM:2005,MaslovTODAES07,SaeediICCAD07,Gupta06,Wille07,WilleDAC09,SaeediMEJ10}).
However, both approaches often lead to sub-optimal circuits with respect to the LNN architectures since the number of elementary gates (i.e.,~quantum cost) are improved without an explicit consideration of the LNN restriction. The same problem exists for quantum circuit synthesis algorithms \cite{mottonen06,Shende06}.

In order to measure the cost of the LNN restriction, a cost metric is defined. Consider a 2-qubit quantum gate $g$ where its control and target are placed at the $c^{th}$ line and at the $t^{th}$ line ($0\leq c,t <n$), respectively. The \emph{NNC (nearest neighbor cost)} of $g$ is defined as $|c-t-1|$ (i.e.,~distance between control and target lines). The NNC of a circuit is defined as the sum of the NNCs of its gates. Optimal NNC for a circuit is 0 where all quantum gates are either 1-qubit or 2-qubit gates performed on adjacent qubits.

\begin{figure}[tb]
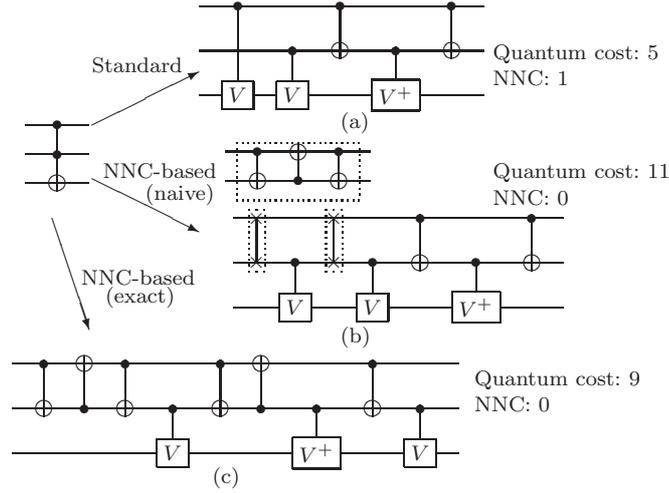

    \centering

\begin{picture}(240,160)

\put(10,120){\input{qcircuits/toffoli.tex}}

\put(35,120){\vector(2,1){40}}
\put(35,140){\footnotesize Standard}
\put(35,100){\vector(2,-1){40}}
\put(39,100){\footnotesize NNC-based}
\put(55,92){\footnotesize (naive)}
\put(20,85){\vector(1,-3){14}}
\put(31,60){\footnotesize NNC-based}
\put(40,53){\footnotesize (exact)}

\put(75,165){\input{qcircuits/toffoli_std_decomp.tex}}
\put(128,119){\footnotesize (a)}
\put(185,145){\footnotesize Quantum cost: 5}
\put(185,135){\footnotesize NNC: 1}

\put(85,110){\input{qcircuits/swap.tex}}
\put(128,38){\footnotesize (b)}
\put(88,85){\input{qcircuits/toffoli_std_decomp_swap.tex}}
\put(185,100){\footnotesize Quantum cost: 11}
\put(185,90){\footnotesize NNC: 0}

\put(5,30){\input{qcircuits/toffoli_exact_decomp.tex}}
\put(80,-15){\footnotesize (c)}
\put(178,22){\footnotesize Quantum cost: 9}
\put(178,12){\footnotesize NNC: 0}

\end{picture}
\vspace{0.5cm}
\caption{Different decompositions of a Toffoli gate}
\label{fig:decomp}
\end{figure}

Since synthesis algorithms may use several non-elementary gates during synthesis, all non-elementary gates should be decomposed into a set of elementary unit-cost gates for physical implementation. Decomposition methods proposed in \cite{Barenco95,MaslovTCAD08,Shende06} are extensively used for this purpose. On the other hand, after applying one of the available synthesis and/or decomposition algorithms, non-optimal circuits with respect to NNC may result. For example, Fig.~\ref{fig:decomp}~(a) shows the standard decomposition of a Toffoli gate which leads to an NNC value of 1. To make this circuit applicable for the LNN architectures, SWAP gates must be applied for each non-adjacent quantum gate. More precisely, SWAP gates are added in front of each gate $g$ with non-adjacent control and target lines to ``move'' the control (target) line of $g$ towards the target (control) line until they become adjacent. Afterwards, SWAP gates are added to restore the original ordering of circuit lines. Similar methods have been applied by previous synthesis methods considering the LNN restriction \cite{Hirata09,mottonen06,Chakrabarti07,Khan08,Shende06,SaeediJETC10}.

\begin{example} \label {example:toffoli}
Consider the standard decomposition of a Toffoli gate as depicted in Fig.~\ref{fig:decomp}~(a). As can be seen, the first gate is non-adjacent. Thus, to achieve NNC-optimality, SWAP gates in front and after the first gate are inserted (see Fig.~\ref{fig:decomp}~(b)). Since each SWAP gate requires 3 elementary quantum gates\footnote{As mentioned above, in certain quantum technologies all two-qubit gates are considered as elementary gates. Thus, in this case the SWAP gate is seen as an elementary gate increasing the costs by~1, instead of~3. While this special case is not considered in the following, it is separately evaluated in the experimental evaluation in Section~\ref{sec:exp}.}, this increases the total quantum cost to~$11$, but leads to an NNC value of 0.
\end{example}

By inserting SWAP gates consecutively for each non-adjacent gate, a quantum circuit with NNC of 0 (and thus applicable to LNN architectures) can be determined in linear time. This method is denoted by \emph{naive NNC-based decomposition} in the rest of this paper. However, as can easily be seen, synthesizing quantum circuits for LNN architectures using this method (or similar approaches like~\cite{Hirata09,mottonen06,Chakrabarti07,Khan08,Shende06}) often leads to a significant increase in the quantum cost.
In contrast, often smaller realizations (with NNC of~$0$) are possible. As an example, consider Fig.~\ref{fig:decomp}~(c) that shows an NNC-optimal decomposition with quantum cost of~$9$ (instead of~$11$).
In the next sections, a synthesis flow is described that explicitly takes NNC into account. Hence, better quantum circuit realizations for the LNN architectures can be found as shown in the experimental results section.

\section {Explicit Consideration of NNC} \label {sec:opt}

In this section, we propose new synthesis and optimization approaches that explicitly take NNC into account. More precisely, a template-matching post-optimization algorithm is introduced to simplify the circuits resulted from the existing synthesis flow. Furthermore, an exact synthesis approach is proposed that determines NNC-optimal circuits with minimal quantum cost. The resulting circuits can later be exploited to optimize large circuits. Finally, two heuristic approaches are introduced that modify the initial qubit locations in order to remove unnecessary SWAP gates and therewith to reduce the cost.

\subsection {NNC-based Template Matching} \label {sec:TM}

The idea of exploiting templates has originally been proposed in~\cite{MillerDAC03} and extended in~\cite{Chakrabarti} for LNN architectures. In this section, further templates for LNN architectures are proposed that outperform the previous ones as shown below.

\begin{figure}
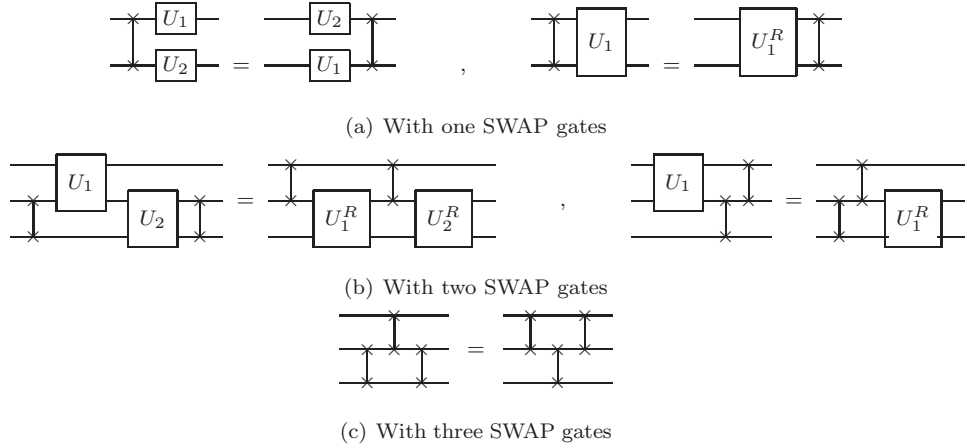

\centering

    \subfigure[With one SWAP gates\label{templ:One-SWAP}]{\input{qcircuits/template1swap.tex}}
\subfigure[With two SWAP gates\label{templ:Two-SWAP}]{\input{qcircuits/template2swap.tex}}
    \subfigure[With three SWAP gates\label{templ:Three-SWAP}]{\input{qcircuits/template3swap.tex}}
  \caption{Proposed templates}
\end{figure}

Two neighboring gates can be interchanged if the target line of the first gate is not equal to the control lines of the second gate and vice versa (\emph{moving rule}). In addition, two neighboring SWAP gates with the same target lines can be removed (\emph{deletion rule}).
The general idea of template matching is to replace a cascade of reversible gates by a different cascade with the same functionality and afterwards applying the moving and deletion rules to optimize the circuit.
By considering this approach, templates with one, two, and three SWAP gates are proposed in Fig.~\ref{templ:One-SWAP}, Fig.~\ref{templ:Two-SWAP}, and Fig.~\ref{templ:Three-SWAP}, respectively.
The $U_i$ boxes thereby represent any one-qubit or two-qubit gate. A $U_i^R$ box represents the same gate as a $U_i$ box, but probably with interchanged control and target lines.

\begin{figure}
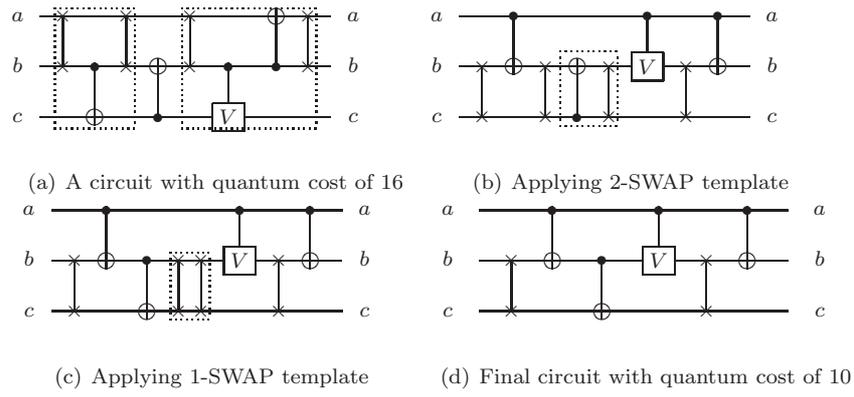

    \centering
    \subfigure[A circuit with quantum cost of 16\label{fig:exam2_temp1}]{
    \input{qcircuits/exam2_temp1.tex}
     }
    \subfigure[Applying 2-SWAP template \label{fig:exam2_temp2}]{
    \input{qcircuits/exam2_temp2.tex}
    }
    \subfigure[Applying 1-SWAP template \label{fig:exam2_temp3}]{
    \input{qcircuits/exam2_temp3.tex}
    }
    \subfigure[Final circuit with quantum cost of 10\label{fig:exam2_temp4}]{
    \input{qcircuits/exam2_temp4.tex}
    }
    \caption{Application of the proposed NNC-based templates} \label{example:temp2}
\end{figure}

As an example, consider the circuit shown in Fig.~\ref{fig:exam2_temp1} with quantum cost of~16. By applying a template introduced in Fig.~\ref{templ:Two-SWAP}, the circuit shown in Fig.~\ref{fig:exam2_temp2} results. Now, a 1-SWAP template (Fig.~\ref{templ:One-SWAP}) can be applied leading to the circuit depicted in Fig.~\ref{fig:exam2_temp3}. Finally, by applying the deletion rule, gates can be removed and, the final quantum cost is improved by about 37\%. The final circuit is shown in Fig.~\ref{fig:exam2_temp4}.

\begin{figure}
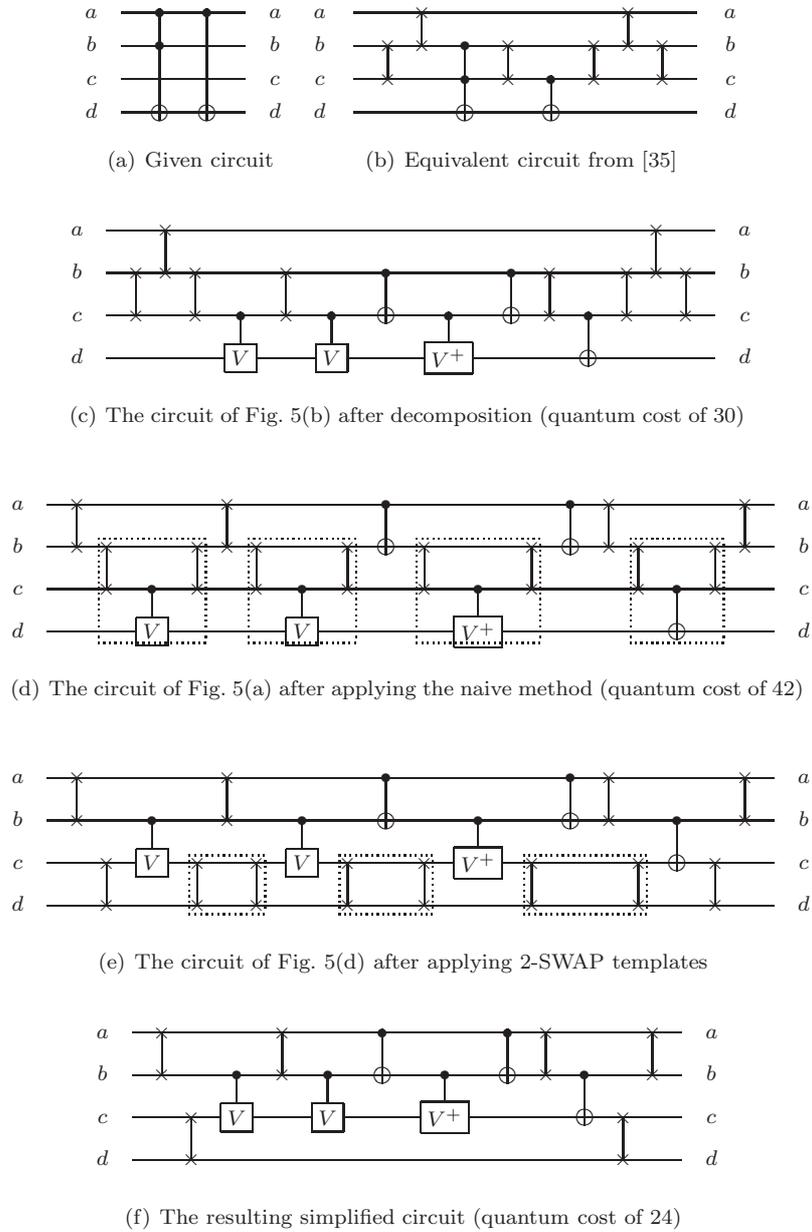

    \centering
    \subfigure[Given circuit \label{fig:exam1_temp1}]{
        \input{qcircuits/exam1_temp1.tex}
    }
    \subfigure[Equivalent circuit from \cite{Chakrabarti}\label{fig:exam1_temp2}]{
        \input{qcircuits/exam1_temp2.tex}
    }
    \subfigure[The circuit of Fig.~\ref{fig:exam1_temp2} after decomposition (quantum cost of~30) \label{fig:exam1_temp3}]{
         \input{qcircuits/exam1_temp3.tex}
   }
    \subfigure[The circuit of Fig.~\ref{fig:exam1_temp1} after applying the naive method (quantum cost of~42) \label{fig:exam1_temp4}]{
          \input{qcircuits/exam1_temp4.tex}
   }
    \subfigure[The circuit of Fig.~\ref{fig:exam1_temp4} after applying 2-SWAP templates \label{fig:exam1_temp5}]{
          \input{qcircuits/exam1_temp5.tex}
    }
    \subfigure[The resulting simplified circuit (quantum cost of 24) \label{fig:exam1_temp6}]{
          \input{qcircuits/exam1_temp6.tex}
    }
    \caption{An existing nearest neighbor template for a Toffoli-CNOT combination and our proposed template}
\end{figure}

The authors of~\cite{Chakrabarti} introduced a set of nearest neighbor templates for Toffoli and CNOT combinations. It can be verified that the introduced templates in Fig.~6(b) and Fig.~6(c) of \cite{Chakrabarti} can be found by applying the deletion rule. Moreover, consider the circuit shown in Fig.~\ref{fig:exam1_temp1} which includes a Toffoli-CNOT combination. Fig.~\ref{fig:exam1_temp2} illustrates a template as proposed in Fig.~6(a) of~\cite{Chakrabarti}. This circuit still has to be decomposed to elementary gates leading to a circuit with quantum cost of 30 as shown in Fig.~\ref{fig:exam1_temp3}. On the other hand, consider the circuit shown in Fig.~\ref{fig:exam1_temp4} obtained by applying the naive method on the circuit of Fig.~\ref{fig:exam1_temp1}. The equivalent circuit after applying the templates introduced in Fig.~\ref{templ:Two-SWAP} is given in Fig.~\ref{fig:exam1_temp5}. Applying the deletion rule finally leads to a circuit with quantum cost of 24 as shown in Fig.~\ref{fig:exam1_temp6}. Thus, applying the templates proposed in this paper in conjunction with the deletion rule improves the result of \cite{Chakrabarti} by 20\%.

Besides that, the efficiency of the proposed templates is illustrated by the following practical relevant example.

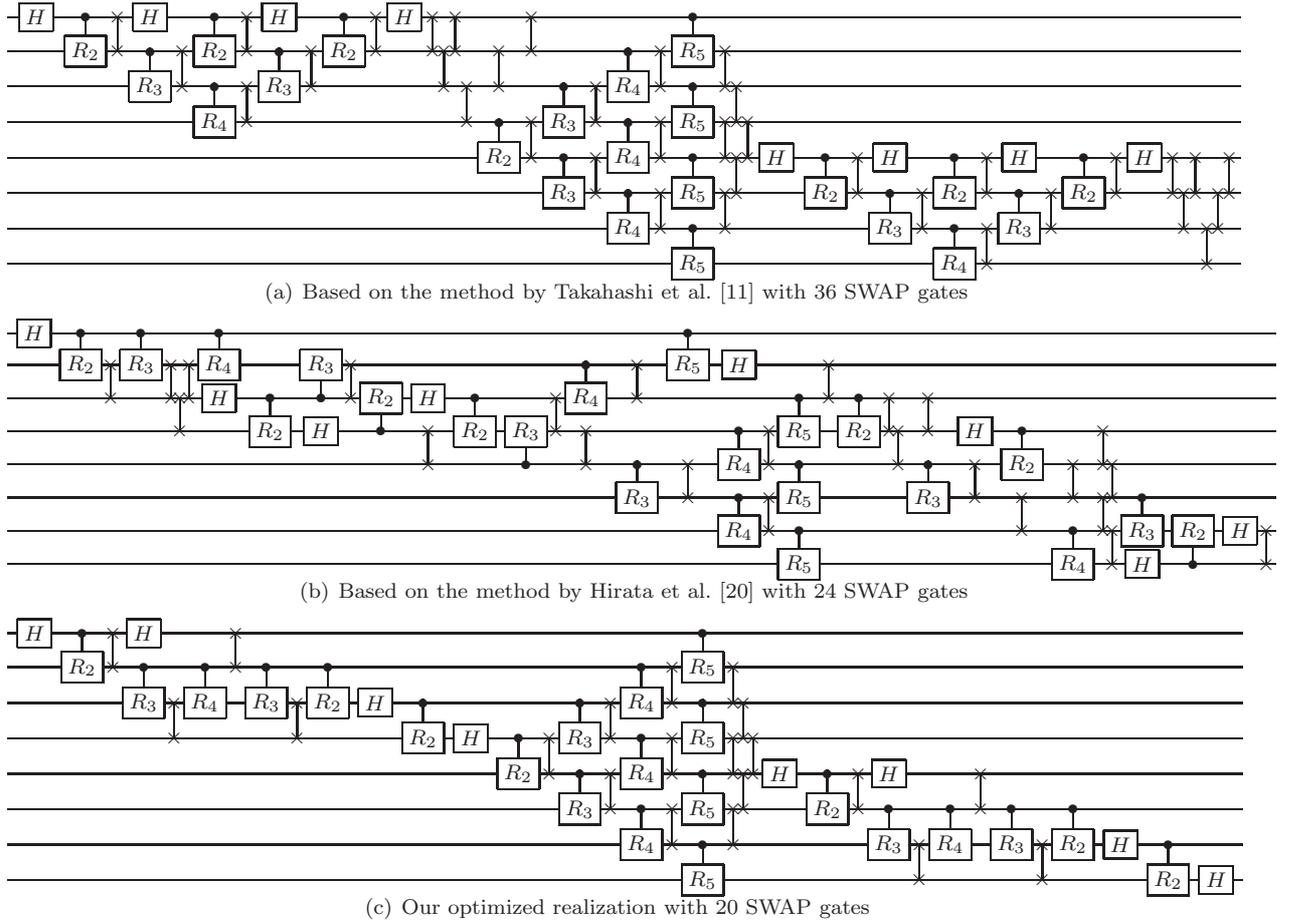
\begin{figure}
\subfigure[Based on the method by Takahashi et al. \cite{Takahashi:2007} with 36 SWAP gates  \label{example:AQFT_org}]{         
\Qcircuit @C=0.5em @R=0.2em {
   & \gate{H} & \ctrl{1}    & \qswap       & \gate{H}   & \qw           & \ctrl{1}  & \qswap        & \gate{H}  & \qw			 & \ctrl{1}    & \qswap     & \gate{H} 		& \qswap				& \qw				& \qswap				& \qw 			& \qw					& \qswap	 & \qw		& \qw		& \qw		 & \qw		& \ctrl{1}	& \qw		 & \qw		& \qw		 & \qw      & \qw         & \qw          & \qw        & \qw           & \qw       & \qw           & \qw 		& \qw        	& \qw         & \qw        & \qw        & \qw          	& \qw        	& \qw           & \qw      	& \qw   	& \qw		& \qw\\
   & \qw      & \gate{R_2}  & \qswap \qwx  & \ctrl{1}   & \qswap        & \gate{R_2}& \qswap \qwx   & \ctrl{1}  & \qswap		& \gate{R_2}  & \qswap \qwx& \qw				& \qswap \qwx 		& \qswap			& \qswap \qwx 		 & \qw 			& \qswap				& \qswap \qwx 	& \qw		 & \qw		& \ctrl{1}	& \qswap	& \gate{R_5}	& \qswap	& \qw		& \qw		& \qw      & \qw         & \qw          & \qw        & \qw           & \qw       & \qw           & \qw 		& \qw        	 & \qw         & \qw        & \qw        & \qw          	& \qw        	& \qw           & \qw      	& \qw           & \qw 		 & \qw\\
   & \qw      & \qw         & \qw          & \gate{R_3} & \qswap \qwx   & \ctrl{1}  & \qswap        & \gate{R_3}& \qswap \qwx	& \qw         & \qw        & \qw 			& \qw 				& \qswap \qwx 	& \qw 				& \qswap			& \qswap \qwx 		& \qw 		& \ctrl{1}	& \qswap	 & \gate{R_4}	& \qswap \qwx	& \ctrl{1}	 & \qswap \qwx	& \qswap	& \qw		 & \qw      & \qw         & \qw          & \qw        & \qw           & \qw       & \qw           & \qw 		& \qw        	& \qw         & \qw        & \qw        & \qw          	& \qw        	 & \qw           & \qw      	& \qw           & \qw 	 & \qw	\\
   & \qw      & \qw         & \qw          & \qw        & \qw           & \gate{R_4}& \qswap \qwx   & \qw       & \qw 			 & \qw         & \qw        & \qw 			& \qw 				& \qw				& \qw 				& \qswap \qwx 	& \ctrl{1}			 & \qswap	& \gate{R_3}	& \qswap \qwx	& \ctrl{1}	& \qswap	& \gate{R_5}	& \qswap	& \qswap \qwx	& \qswap	& \qw      & \qw         & \qw          & \qw        & \qw           & \qw       & \qw           & \qw 		& \qw        	& \qw         & \qw        & \qw        & \qw          	& \qw        	 & \qw           & \qw      	& \qw           & \qw 	& \qw	 \\
   & \qw      & \qw         & \qw          & \qw        & \qw           & \qw       & \qw           & \qw       & \qw        	 & \qw         & \qw        & \qw        	& \qw          	& \qw        	& \qw           	& \qw      		& \gate{R_2}   	& \qswap \qwx 	& \ctrl{1}	 & \qswap	& \gate{R_4}	& \qswap \qwx	& \ctrl{1}	& \qswap \qwx	 & \qswap	& \qswap \qwx	& \gate{H} & \ctrl{1}    & \qswap       & \gate{H}   & \qw           & \ctrl{1}  & \qswap        & \gate{H}  	& \qw		& \ctrl{1}    & \qswap     & \gate{H} 	& \qswap	& \qw		& \qswap	& \qw 		& \qw		 & \qswap	& \qw\\
   & \qw      & \qw         & \qw          & \qw        & \qw           & \qw       & \qw           & \qw       & \qw        	 & \qw         & \qw        & \qw        	& \qw          	& \qw        	& \qw           	& \qw      		& \qw           	 & \qw 		& \gate{R_3}	 & \qswap \qwx	 & \ctrl{1}	& \qswap	& \gate{R_5}	& \qswap	& \qswap \qwx	& \qw		& \qw      & \gate{R_2}  & \qswap \qwx  & \ctrl{1}   & \qswap        & \gate{R_2}& \qswap \qwx   & \ctrl{1}  	 & \qswap	& \gate{R_2}  & \qswap \qwx& \qw	& \qswap \qwx 	& \qswap	& \qswap \qwx 	& \qw 		& \qswap	& \qswap \qwx & \qw	\\
   & \qw      & \qw         & \qw          & \qw        & \qw           & \qw       & \qw           & \qw       & \qw        	 & \qw         & \qw        & \qw        	& \qw          	& \qw        	& \qw           	& \qw      		& \qw           	 & \qw 		& \qw           & \qw 		& \gate{R_4}	& \qswap \qwx	& \ctrl{1}	& \qswap \qwx	& \qw		& \qw		& \qw      & \qw         & \qw          & \gate{R_3} & \qswap \qwx   & \ctrl{1}  & \qswap        & \gate{R_3}	& \qswap \qwx	& \qw         & \qw        & \qw 	& \qw 		 & \qswap \qwx 	& \qw 		& \qswap	 & \qswap \qwx 	& \qw & \qw		\\
   & \qw      & \qw         & \qw          & \qw        & \qw           & \qw       & \qw           & \qw       & \qw        	 & \qw         & \qw        & \qw        	& \qw          	& \qw        	& \qw           	& \qw      		& \qw           	 & \qw 		& \qw           & \qw 		& \qw           & \qw 		& \gate{R_5}	& \qw		& \qw		& \qw		 & \qw      & \qw         & \qw          & \qw        & \qw           & \gate{R_4}& \qswap \qwx   & \qw       	& \qw 		& \qw         & \qw        & \qw 	& \qw 		 & \qw		& \qw 		& \qswap \qwx 	& \qw		 & \qw  & \qw \\
   }
}
\subfigure[Based on the method by Hirata et al. \cite{Hirata09} with 24 SWAP gates  \label{example:AQFT_Hirata}]{          
\Qcircuit @C=0.40em @R=0.1em {
& \gate{H}	& \ctrl{1}	& \qw		& \ctrl{1}	& \qw		& \qw		& \qw		& \ctrl{1}	& \qw		& \qw		 & \qw		 & \qw		& \qw		 & \qw		& \qw		& \qw		& \qw		& \qw		& \ctrl{1}	& \qw		 & \qw		& \qw		 & \qw		 & \qw		& \qw		& \qw		 & \qw		& \qw		& \qw		& \qw		 & \qw		& \qw		& \qw		 & \qw		& \qw		& \qw	& \qw	\\
& \qw		& \gate{R_2}	& \qswap	& \gate{R_3}	& \qswap	& \qw		& \qswap	& \gate{R_4}	& \qw		 & \gate{R_3}	 & \qswap	& \qw		& \qw		& \qw		& \qw		& \qw		& \ctrl{1}	& \qswap	& \gate{R_5}	& \gate{H}	& \qw		 & \qw		& \qswap	& \qw		 & \qw		& \qw		& \qw		& \qw		 & \qw		& \qw		 & \qw		& \qw		& \qw		& \qw		& \qw		& \qw	& \qw	 \\
& \qw		& \qw		& \qswap \qwx	& \qw		& \qswap \qwx	& \qswap	& \qswap \qwx	& \gate{H}	& \ctrl{1}	 & \ctrl{-1}	& \qswap \qwx	 & \gate{R_2}	& \gate{H}	& \ctrl{1}	& \qw		& \qswap	& \gate{R_4}	& \qswap \qwx	& \qw		 & \qw		& \qw		& \ctrl{1}	& \qswap \qwx	& \ctrl{1}	& \qswap	& \qw		& \qswap	 & \qw		& \qw		& \qw		 & \qw		& \qw		& \qw		& \qw		& \qw		 & \qw	& \qw	\\
& \qw		& \qw		& \qw		& \qw		& \qw		& \qswap \qwx	& \qw		& \qw		& \gate{R_2}	& \gate{H}	& \qw		 & \ctrl{-1}	& \qswap	& \gate{R_2}	& \gate{R_3}	& \qswap \qwx	& \qswap	& \qw		 & \qw		& \ctrl{1}	 & \qswap	& \gate{R_5}	& \qw		& \gate{R_2}	& \qswap \qwx	& \qswap	& \qswap \qwx	& \gate{H}	& \ctrl{1}	& \qw		 & \qswap	& \qw		& \qw		& \qw		 & \qw		& \qw		& \qw\\
& \qw		& \qw		& \qw		& \qw		& \qw		& \qw		& \qw		& \qw		& \qw		& \qw		 & \qw		 & \qw		& \qswap \qwx	& \qw		& \ctrl{-1}	& \qw		& \qswap \qwx	& \ctrl{1}	& \qswap	& \gate{R_4}	& \qswap \qwx	& \ctrl{1}	& \qw		& \qw		 & \qw		& \qswap \qwx	& \ctrl{1}	& \qswap	& \gate{R_2}	& \qswap	& \qswap \qwx	& \qswap	& \qw		& \qw		& \qw		& \qw	 & \qw	\\
& \qw		& \qw		& \qw		& \qw		& \qw		& \qw		& \qw		& \qw		& \qw		& \qw		 & \qw		 & \qw		& \qw		 & \qw		& \qw		& \qw		& \qw		& \gate{R_3}	& \qswap \qwx	& \ctrl{1}	& \qswap	& \gate{R_5}	& \qw		& \qw		& \qw		 & \qw		& \gate{R_3}	& \qswap \qwx	 & \qswap	& \qswap \qwx	& \qswap	 & \qswap \qwx	& \ctrl{1}	& \qw		& \qw		& \qw		& \qw\\
& \qw		& \qw		& \qw		& \qw		& \qw		& \qw		& \qw		& \qw		& \qw		& \qw		 & \qw		 & \qw		& \qw		 & \qw		& \qw		& \qw		& \qw		& \qw		& \qw		& \gate{R_4}	& \qswap \qwx	& \ctrl{1}	 & \qw		& \qw		& \qw		 & \qw		& \qw		& \qw		& \qswap \qwx	& \ctrl{1}	& \qswap \qwx	& \qswap	& \gate{R_3}	& \gate{R_2}	& \gate{H}	& \qswap	& \qw\\
& \qw		& \qw		& \qw		& \qw		& \qw		& \qw		& \qw		& \qw		& \qw		& \qw		 & \qw		 & \qw		& \qw		 & \qw		& \qw		& \qw		& \qw		& \qw		& \qw		& \qw		 & \qw		& \gate{R_5}	& \qw		& \qw		& \qw		& \qw		& \qw		& \qw		& \qw		& \gate{R_4}	& \qw		& \qswap \qwx	& \gate{H}	 & \ctrl{-1}	& \qw		& \qswap \qwx	& \qw\\
}
}
\subfigure[Our optimized realization with 20 SWAP gates  \label{example:AQFT_Ours}]{ 
\Qcircuit @C=0.43em @R=0.2em {
   & \gate{H} & \ctrl{1}    & \qswap       & \gate{H}   & \qw           & \qw  		& \qswap        & \qw	  	& \qw		 & \qw          	 & \qw          	 & \qw 		& \qw          	& \qw 		& \qw		& \qw		& \qw		& \qw		 & \qw		& \ctrl{1}	 & \qw		& \qw		& \qw		 & \qw      & \qw         & \qw          & \qw        & \qw           & \qw       	& \qw           & \qw       	& \qw        	& \qw        	 & \qw         	& \qw     	 & \qw     & \qw 	\\
   & \qw      & \gate{R_2}  & \qswap \qwx  & \ctrl{1}   & \qw	        & \ctrl{1}  	& \qswap \qwx   & \ctrl{1}	& \qw		& \ctrl{1}    	& \qw          	 & \qw 		& \qw          	& \qw 		& \qw 		& \qw		& \qw		 & \ctrl{1}	& \qswap	 & \gate{R_5}	& \qswap	& \qw		& \qw		 & \qw      & \qw         & \qw          & \qw        & \qw           & \qw       	 & \qw           & \qw       	& \qw        	& \qw        	 & \qw         	 & \qw          	& \qw        & \qw   	\\
   & \qw      & \qw         & \qw          & \gate{R_3} & \qswap        & \gate{R_4}	& \qw		& \gate{R_3}	& \qswap        & \gate{R_2}	& \gate{H}     	& \ctrl{1}	& \qw          	& \qw 		& \qw 		& \ctrl{1}	& \qswap	& \gate{R_4}	& \qswap \qwx	& \ctrl{1}	& \qswap \qwx	& \qswap	& \qw		& \qw      & \qw         & \qw          & \qw        & \qw           & \qw       	 & \qw           & \qw       	& \qw        	& \qw        	 & \qw         	& \qw          	& \qw        & \qw   	 \\
   & \qw      & \qw         & \qw          & \qw        & \qswap \qwx   & \qw       	& \qw       	& \qw       	 & \qswap \qwx   & \qw     	& \qw          	& \gate{R_2}   	& \gate{H}     	& \ctrl{1}	& \qswap	& \gate{R_3}	 & \qswap \qwx	& \ctrl{1}	& \qswap	& \gate{R_5}	& \qswap	& \qswap \qwx	& \qswap	& \qw      & \qw         & \qw          & \qw        & \qw           & \qw       	& \qw           & \qw       	 & \qw        	& \qw        	 & \qw         	& \qw          	& \qw       & \qw    	\\
   & \qw      & \qw         & \qw          & \qw        & \qw           & \qw       	& \qw           & \qw       	 & \qw        	 & \qw     	& \qw     	& \qw 		& \qw          	& \gate{R_2}   	& \qswap \qwx 	& \ctrl{1}	 & \qswap	& \gate{R_4}	& \qswap \qwx	& \ctrl{1}	& \qswap \qwx	& \qswap	& \qswap \qwx	& \gate{H} & \ctrl{1}    & \qswap       & \gate{H}   & \qw           & \qw  		& \qswap        & \qw	  	& \qw		& \qw          	 & \qw          	& \qw 		& \qw        & \qw   	 \\
   & \qw      & \qw         & \qw          & \qw        & \qw           & \qw       	& \qw           & \qw       	 & \qw        	 & \qw          	 & \qw          	& \qw           & \qw          	& \qw           & \qw 		& \gate{R_3}	 & \qswap \qwx	& \ctrl{1}	& \qswap	& \gate{R_5}	 & \qswap	& \qswap \qwx	& \qw		& \qw      & \gate{R_2}  & \qswap \qwx  & \ctrl{1}   & \qw	     & \ctrl{1}  	& \qswap \qwx   & \ctrl{1}	& \qw		& \ctrl{1}    	& \qw          	& \qw 		& \qw          & \qw 	\\
   & \qw      & \qw         & \qw          & \qw        & \qw           & \qw       	& \qw           & \qw       	 & \qw        	 & \qw          	 & \qw          	& \qw           & \qw          	& \qw           & \qw 		& \qw           & \qw 		& \gate{R_4}	& \qswap \qwx	& \ctrl{1}	 & \qswap \qwx	& \qw		& \qw		& \qw      & \qw         & \qw          & \gate{R_3} & \qswap        & \gate{R_4}	& \qw		& \gate{R_3}	& \qswap        & \gate{R_2}	& \gate{H}     	& \ctrl{1}	 & \qw    & \qw       	\\
   & \qw      & \qw         & \qw          & \qw        & \qw           & \qw       	& \qw           & \qw       	 & \qw        	 & \qw          	 & \qw          	& \qw           & \qw          	& \qw           & \qw 		& \qw           & \qw 		& \qw           & \qw 		& \gate{R_5}	 & \qw		& \qw		& \qw		& \qw      & \qw         & \qw          & \qw        & \qswap \qwx   & \qw       	& \qw       	& \qw       	 & \qswap \qwx   & \qw     	& \qw          	& \gate{R_2}   	& \gate{H}     & \qw 	\\
}
}
\caption{Circuits realizing the Approximate Quantum Fourier Transform (AQFT)}
\end{figure}

\begin{example}\label{example:template3}
Consider the circuit shown in Fig.~\ref{example:AQFT_org} which is the approximate quantum Fourier transform circuit (AQFT) \cite{PhysRevA.54.139}
with 36 SWAP gates
obtained by the method of~\cite{Takahashi:2007} for $8$ qubits and an approximation parameter of $5$. Note that $R_k$ is the rotation by $2\pi/2^k$ and $H$ is the Hadamard gate. Fig.~\ref{example:AQFT_Hirata} is an equivalent circuit
with 24 SWAP gates
constructed by a method recently introduced in~\cite{Hirata09}. On the other hand, applying the proposed templates on the result of~\cite{Takahashi:2007} leads to the circuit
with 20 SWAP gates
illustrated in Fig.~\ref{example:AQFT_Ours}.
\end{example}

\subsection{Exploiting Exact Synthesis}\label {sec:opt_exact}
A few exact synthesis methods for quantum circuits have recently been introduced. They generate quantum circuits with minimal quantum cost (for examples see \cite {Hung06,GrosseMVL08}). However, no approach to determine optimal circuits for LNN architectures has been proposed so far. In this section, an exact synthesis algorithm is proposed to construct quantum circuits with \emph{both}, minimal quantum cost and minimal NNC.

The developed approach is similar to the one introduced in~\cite{GrosseMVL08}.
Here, the synthesis problem is expressed as a sequence of \emph{Boolean satisfiability} (SAT) instances.
For a given function~$f$, it is checked if a circuit with~$c$ gates realizing~$f$ exists. Thereby,~$c$ is initially assigned to~$1$ and increased in each iteration if no realization is found.

More formally, for a given $c$ and a reversible function $f:\mathbb{B}^n \rightarrow \mathbb{B}^n$, the following SAT instance is created:

$$\Phi \wedge \bigwedge\limits_{i=0}^{2^n-1}([\overrightarrow{inp}_i]_2 = i \wedge [\overrightarrow{out}_i]_2 = f(i))\mbox{,}$$
where
\begin{itemize}
\item $\overrightarrow{inp}_i$ is a Boolean vector representing the inputs of the network to be synthesized for
truth table line~$i$,
\item $\overrightarrow{out}_i$ is a Boolean vector representing the outputs of the network to be synthesized for
truth table line~$i$, and
\item $\Phi$ is a set of constraints representing the synthesis problem for a given gate library.
\end{itemize}
The difference in comparison to~\cite{GrosseMVL08} is that the constraints in $\Phi$ do not represent the whole set of elementary quantum gates and a restricted gate library with only adjacent gates is applied.

Although solving the generated SAT instances using a modern SAT solver can produce optimized circuits, the applicability of the exact method is limited to functions with a small number of qubits and gates due to the exponential search space. Actually, the proposed exact method is sufficient to construct minimal realizations with respect to both quantum cost and NNC for a set of Toffoli and Peres gate configurations as shown in Table~\ref {tab:MacrosTable}. However, these optimal circuits can be exploited to improve the naive NNC-based decomposition method. More precisely, once an exact NNC-optimal quantum circuit for a function is available (denoted by \emph{macro} in the following), the decomposition from the naive approach is replaced by the optimal circuit. The following example illustrates the idea. 

\begin{table}[!t]
	\caption{List of available macros}
	\label{tab:MacrosTable}
	\centering
		\begin{tabular}{|c|c|c|c|c|}
			\hline
			\multirow{2}{*}{$n$} & \multirow{2}{*} {Macro} & \multicolumn{ 2}{|c|}{Cost} & \multirow{2}{*}{Impr.}\\
			    &       &  Naive &Exact & \\
			\hline
			3 & P(\{a,b\},c), P(\{c,b\},a) & 12 & 8 & 33\%\\
			3 & P(\{a,c\},b), P(\{c,a\},b) & 24& 12  & 50\%\\
			4 & P(\{a,b\},d), P(\{d,c\},a) & 30 & 11 & 63\%\\
			3 & t2(\{a,b\},c), t2(\{c,b\},a) & 11 & 9 & 18\%\\
			4 & t2(\{a,b\},d), t2(\{d,c\},a) & 29 & 12 & 59\%\\
			3 & t2(\{a,c\},b) & 17 & 13 & 24\%\\
			4 & t2(\{d,b\},a), t2(\{a,c\},d) & 29 &13 & 55\%\\
			\hline
		\end{tabular}
\end{table}

\begin{figure}
           \input{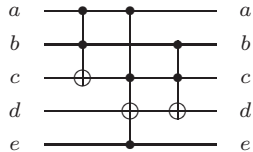}
 \caption{Circuit of Example \ref{exmacro} \label{fig:example_circuit}  }
\end{figure}

\begin{figure}
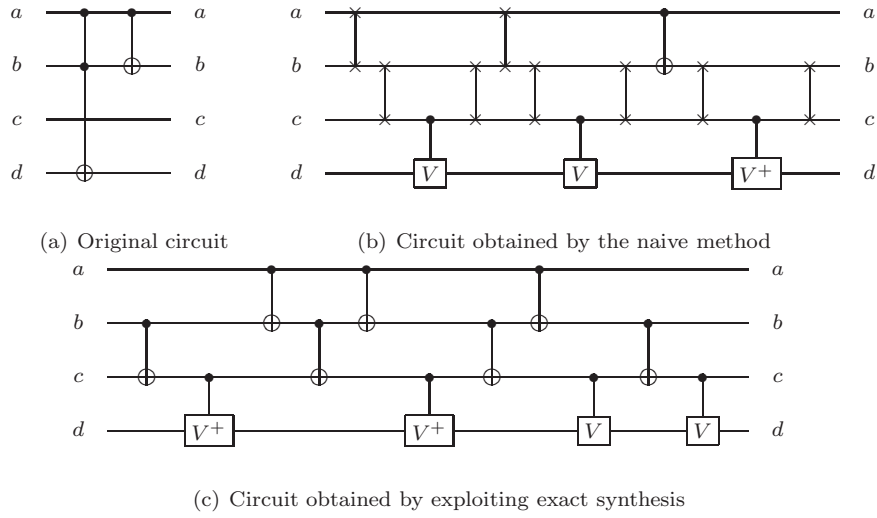

        \centering
        \subfigure[Original circuit\label{fig:Peres}]{
           \input{qcircuits/4_p_abd.tex}
        }\subfigure[Circuit obtained by the naive method\label{fig:PeresNaive}]{
           \input{qcircuits/4_p_abd_naive.tex}
        } \subfigure[Circuit obtained by exploiting exact synthesis\label{fig:PeresExact}]{
           \input{qcircuits/4_p_abd_eq.tex}
         }
    	\caption{NNC-based synthesis of a Peres gate}     	\label{fig:exactsample}
\end{figure}

\begin{example} \label {exmacro}
Reconsider the decomposition of a Toffoli gate as depicted in Fig.~\ref{fig:decomp}. By applying the proposed exact synthesis approach, an NNC-optimal quantum circuit as shown in Fig.~\ref{fig:decomp}(c) results. In comparison to the naive method (see Fig.~\ref{fig:decomp}(b)), this reduces the quantum cost from 11 to 9 while still ensuring NNC optimality.

After finding the optimal decomposition of a given gate, it can be used as a macro to simplify other circuits. For example, consider the circuit shown in Fig.~\ref{fig:example_circuit}. Here, for the second gate the naive method is applied and SWAPs are added, while for the remaining ones the obtained macro is used. This enables a quantum cost reduction from 96 to 92.

Moreover, Fig.~\ref{fig:PeresNaive} and Fig.~\ref{fig:PeresExact} show the NNC-optimal circuit of the Peres gate obtained by the naive and by the exact approach, respectively. As illustrated, applying the naive approach leads to quantum cost of 28
while the optimal circuit has only quantum cost of 11.
\end{example}

In total, we generated 13 macros as listed in Table \ref {tab:MacrosTable} together with the respective costs in comparison to the costs obtained by using the naive method. As can be seen, exploiting these macros reduces the cost for each gate by up to 63\%. The effect of these macros on the decomposition of larger circuits is considered in the experimental results section in detail.

\subsection{Reordering Circuit Lines}\label {sec:opt_reordering}
\begin{sidewaystable}[t]
\vspace{+16cm}
\scriptsize
\centering
\caption{Experimental results (considering a SWAP gate to be composed out of 3 elementary gates)}\label{table:nnc}
\begin{tabular}{||l||r|r|r|r|r|r|r|r|r|r|r|r|r|r|r|r|r||r|r|r|r|r|r|r|r|r|r|r|r|r|r|r|r|r||}
\hline
 \multirow{3}{*}{\textsc{circuit}}  & \multicolumn{4}{|c|}{Original circuit} & \multicolumn{11}{|c|}{Decomposed circuit} & \multirow{2}{*}{\textsc{Time}} & \multirow{3}{*}{\textsc{Ohead}}\\
 &  &  &  &  &  & \multicolumn{1}{|c|}{N} & \multicolumn{1}{|c|}{T} & \multicolumn{1}{|c|}{M} & \multicolumn{1}{|c|}{MT} & \multicolumn{1}{|c|}{G} & \multicolumn{1}{|c|}{L} & \multicolumn{1}{|c|}{GL} & \multicolumn{2}{|c|}{Best} & \multicolumn{1}{|c|}{Best} &  & \\
 & $n$  & gc  & qc  & NNC  & $n$  & qc  & qc  & qc  & qc  & qc  & qc  & qc  & qc  & Method  & Impr.\%  &  Sec. &  \\
\hline
\hline
 0410184\_169  & 14  & 46  & 90  & 68  & 14  & 234  & 212  & 197  & 189  & 234  & 423  & 423  & 189  & MT  & 19  & 0  & 2.10 \\
 3\_17\_13  & 3  & 6  & 14  & 8  & 3  & 32  & 32  & 28  & 26  & 32  & 32  & 32  & 26  & MT  & 18  & 0  & 1.86 \\
 4\_49\_17  & 4  & 12  & 32  & 64  & 4  & 158  & 104  & 120  & 102  & 128  & 98  & 98  & 92  & GT  & 41  & 0  & 2.88 \\
 4gt10-v1\_81  & 5  & 6  & 34  & 164  & 5  & 282  & 120  & 282  & 120  & 258  & 150  & 147  & 120  & T  & 57  & 0  & 3.53 \\
 4gt11\_84  & 5  & 3  & 7  & 26  & 5  & 49  & 31  & 47  & 29  & 25  & 22  & 16  & 14  & MGL  & 71  & 0  & 2.00 \\
 4gt12-v1\_89  & 5  & 5  & 42  & 320  & 6  & 525  & 195  & 525  & 195  & 321  & 171  & 168  & 141  & GT  & 73  & 0  & 3.36 \\
 4gt13-v1\_93  & 5  & 4  & 16  & 104  & 5  & 173  & 83  & 173  & 83  & 77  & 56  & 53  & 47  & GT  & 72  & 0  & 2.94 \\
 4gt4-v0\_80  & 5  & 5  & 34  & 218  & 6  & 366  & 168  & 364  & 166  & 168  & 138  & 141  & 132  & GT  & 63  & 0  & 3.88 \\
 4gt5\_75  & 5  & 5  & 21  & 76  & 5  & 142  & 94  & 138  & 96  & 118  & 82  & 79  & 70  & GT  & 50  & 0  & 3.33 \\
 4mod5-v1\_23  & 5  & 8  & 24  & 90  & 5  & 174  & 84  & 155  & 101  & 114  & 78  & 78  & 72  & GLT  & 58  & 0  & 3.00 \\
 4mod7-v0\_95  & 5  & 6  & 38  & 144  & 5  & 256  & 140  & 256  & 140  & 352  & 127  & 121  & 121  & GL  & 52  & 0  & 3.18 \\
 add16\_174  & 49  & 64  & 192  & 220  & 49  & 762  & 446  & 473  & 473  & 762  & 1104  & 1104  & 428  & MGL  & 43  & 0  & 2.23 \\
 add32\_183  & 97  & 128  & 384  & 444  & 97  & 1530  & 894  & 953  & 953  & 1530  & 3744  & 3744  & 860  & MGL  & 43  & 2  & 2.24 \\
 add64\_184  & 193  & 256  & 768  & 892  & 193  & 3066  & 1790  & 1913  & 1913  & 3066  & 13632  & 13632  & 1724  & MGL  & 43  & 14  & 2.24 \\
 add8\_172  & 25  & 32  & 96  & 108  & 25  & 378  & 222  & 233  & 233  & 378  & 360  & 360  & 212  & MGL  & 43  & 0  & 2.21 \\
 aj-e11\_165  & 4  & 13  & 45  & 144  & 5  & 280  & 166  & 260  & 164  & 280  & 181  & 181  & 160  & GT  & 42  & 0  & 3.56 \\
 alu-v4\_36  & 5  & 7  & 31  & 136  & 5  & 242  & 146  & 238  & 148  & 218  & 113  & 104  & 98  & GT  & 59  & 0  & 3.16 \\
 cnt3-5\_180  & 16  & 20  & 120  & 1634  & 16  & 2621  & 613  & 2591  & 677  & 1457  & 731  & 728  & 511  & GT  & 80  & 0  & 4.26 \\
 cycle10\_2\_110  & 12  & 19  & 1126  & 13472  & 12  & 21420  & 13700  & 21420  & 13700  & 21420  & 8046  & 8046  & 7874  & LT  & 63  & 4  & 6.99 \\
 decod24-v3\_46  & 4  & 9  & 9  & 36  & 4  & 63  & 27  & 63  & 27  & 39  & 21  & 24  & 21  & L  & 66  & 0  & 2.33 \\
 ham15\_108  & 15  & 70  & 453  & 9978  & 15  & 15494  & 10610  & 15390  & 10582  & 14030  & 2627  & 2588  & 2588  & GL  & 83  & 2  & 5.71 \\
 ham7\_104  & 7  & 23  & 83  & 624  & 7  & 1035  & 681  & 1027  & 695  & 657  & 342  & 333  & 327  & GT  & 68  & 0  & 3.94 \\
 hwb4\_52  & 4  & 11  & 23  & 40  & 4  & 107  & 77  & 83  & 63  & 107  & 65  & 65  & 63  & MT  & 41  & 0  & 2.74 \\
 hwb5\_55  & 5  & 24  & 104  & 470  & 5  & 823  & 407  & 817  & 415  & 595  & 337  & 340  & 335  & LT  & 59  & 0  & 3.22 \\
 hwb6\_58  & 6  & 42  & 142  & 710  & 6  & 1304  & 692  & 1160  & 672  & 1268  & 614  & 545  & 542  & GLT  & 58  & 0  & 3.82 \\
 hwb7\_62  & 7  & 331  & 2325  & 16890  & 8  & 27967  & 15547  & 27869  & 15533  & 25939  & 13390  & 12955  & 12853  & LT  & 54  & 4  & 5.53 \\
 hwb8\_118  & 8  & 633  & 14260  & 115030  & 9  & 187272  & 96906  & 186880  & 96834  & 182196  & 87495  & 87498  & 87495  & L  & 53  & 39  & 6.14 \\
 hwb9\_123  & 9  & 1959  & 18124  & 189426  & 10  & 304659  & 168147  & 304540  & 168160  & 302481  & 124068  & 124041  & 124041  & GL  & 59  & 74  & 6.84 \\
 mod5adder\_128  & 6  & 15  & 83  & 600  & 6  & 1011  & 435  & 978  & 432  & 675  & 330  & 333  & 330  & L  & 67  & 0  & 3.98 \\
 mod8-10\_177  & 5  & 14  & 88  & 582  & 6  & 975  & 407  & 969  & 409  & 621  & 372  & 363  & 317  & GT  & 67  & 0  & 3.60 \\
 plus127mod8192\_162  & 13  & 910  & 57400  & 661596  & 14  & 1057946  & 675624  & 1057804  & 675610  & 1057946  & 503516  & 503516  & 496698  & LT  & 53  & 376  & 8.65 \\
 plus63mod4096\_163  & 12  & 429  & 25492  & 254864  & 13  & 407926  & 256792  & 407784  & 256778  & 407926  & 210400  & 210400  & 210100  & LT  & 48  & 113  & 8.24 \\
 plus63mod8192\_164  & 13  & 492  & 32578  & 397864  & 14  & 633994  & 409384  & 633852  & 409358  & 633994  & 279016  & 279016  & 271030  & LT  & 57  & 187  & 8.32 \\
 rd32-v0\_67  & 4  & 2  & 10  & 10  & 4  & 38  & 26  & 19  & 17  & 20  & 32  & 20  & 17  & MT  & 55  & 0  & 1.70 \\
 rd53\_135  & 7  & 16  & 77  & 466  & 7  & 822  & 450  & 750  & 456  & 702  & 330  & 303  & 303  & GL  & 63  & 0  & 3.94 \\
 rd73\_140  & 10  & 20  & 76  & 450  & 10  & 790  & 400  & 739  & 401  & 646  & 304  & 295  & 286  & LT  & 63  & 0  & 3.76 \\
 rd84\_142  & 15  & 28  & 112  & 910  & 15  & 1516  & 626  & 1465  & 639  & 1696  & 556  & 586  & 556  & L  & 63  & 0  & 4.96 \\
 sym9\_148  & 10  & 210  & 4368  & 48736  & 10  & 77556  & 46110  & 77556  & 46110  & 67428  & 20643  & 25023  & 20640  & LT  & 73  & 11  & 4.73 \\
 sys6-v0\_144  & 10  & 15  & 67  & 358  & 10  & 638  & 326  & 587  & 329  & 842  & 263  & 308  & 263  & L  & 58  & 0  & 3.93 \\
 urf1\_149  & 9  & 11554  & 57770  & 462708  & 9  & 794582  & 353732  & 735170  & 329762  & 659150  & 238475  & 238490  & 238475  & L  & 69  & 261  & 4.13 \\
 urf2\_152  & 8  & 5030  & 25150  & 171284  & 8  & 297178  & 133606  & 276882  & 126348  & 297178  & 101683  & 101683  & 101656  & LT  & 65  & 56  & 4.04 \\
 urf3\_155  & 10  & 26468  & 132340  & 1282724  & 10  & 2121808  & 897358  & 2038584  & 874848  & 1933372  & 596368  & 596371  & 596356  & LT  & 71  & 1298  & 4.51 \\
 urf5\_158  & 9  & 10276  & 51380  & 442748  & 9  & 740084  & 333674  & 706412  & 321496  & 667484  & 208709  & 208706  & 208700  & LT  & 71  & 231  & 4.06 \\
 urf6\_160  & 15  & 10740  & 53700  & 951276  & 15  & 1487904  & 589662  & 1478080  & 586572  & 1334916  & 320412  & 320409  & 320400  & LT  & 78  & 596  & 5.97 \\
\hline
\end{tabular}
\textsc{N}: Naive method (i.e., synthesis, decomposition, SWAP insertion)\hspace{.3cm}
\textsc{T}: With template matching \hspace{.3cm}
\textsc{M}: With macros replacement \hspace{.3cm}
\textsc{G}: With global reordering\hspace{.3cm}
\textsc{L}: With local reordering \hspace{.3cm}

\begin{flushleft}
Column \emph{Time} denotes the overall run-time needed to generate the results for \emph{all} possible configurations.\\
Almost all results have been obtained in negligible run-time (i.e., in less than one CPU second). Only if template matching was enabled more run-time was needed.

\end{flushleft}

\end{sidewaystable}

\begin{sidewaystable}[t]
\vspace{+16cm}
\scriptsize
\centering
\caption{Experimental results (considering a SWAP gate as an elementary gate)}\label{table:nnc2}
\begin{tabular}{||l||r|r|r|r|r|r|r|r|r|r|r|r|r|r|r|r|r||r|r|r|r|r|r|r|r|r|r|r|r|r|r|r|r|r||}
\hline
 circuit  & \multicolumn{4}{|c|}{Original circuit} & \multicolumn{11}{|c|}{Decomposed circuit} & Time & Overhead\\
 &  &  &  &  &  & \multicolumn{1}{|c|}{N} & \multicolumn{1}{|c|}{NT} & \multicolumn{1}{|c|}{NM} & \multicolumn{1}{|c|}{NMT} & \multicolumn{1}{|c|}{G} & \multicolumn{1}{|c|}{L} & \multicolumn{1}{|c|}{GL} & \multicolumn{2}{|c|}{Best} & \multicolumn{1}{|c|}{Best} & Sec. & \\
 & $n$  & gc  & qc  & NNC  & $n$  & qc  & qc  & qc  & qc  & qc  & qc  & qc  & qc  & Method  & Impr.\%  &   &  \\
\hline
\hline
 0410184\_169  & 14  & 46  & 90  & 68  & 14  & 138  & 124  & 157  & 149  & 138  & 201  & 201  & 124  & NT  & 10  & 0.10  & 1.38 \\
 3\_17\_13  & 3  & 6  & 14  & 8  & 3  & 20  & 20  & 24  & 22  & 20  & 20  & 20  & 20  & N  & 0  & 0.01  & 1.43 \\
 4\_49\_17  & 4  & 12  & 32  & 64  & 4  & 74  & 56  & 76  & 70  & 64  & 54  & 54  & 52  & GT  & 29  & 0.01  & 1.62 \\
 4gt10-v1\_81  & 5  & 6  & 34  & 164  & 5  & 118  & 64  & 118  & 64  & 110  & 74  & 73  & 64  & NT  & 45  & 0.02  & 1.88 \\
 4gt11\_84  & 5  & 3  & 7  & 26  & 5  & 21  & 15  & 23  & 17  & 13  & 12  & 10  & 10  & GL  & 52  & 0.00  & 1.43 \\
 4gt12-v1\_89  & 5  & 5  & 42  & 320  & 6  & 205  & 95  & 205  & 95  & 137  & 87  & 86  & 77  & GT  & 62  & 0.03  & 1.83 \\
 4gt13-v1\_93  & 5  & 4  & 16  & 104  & 5  & 69  & 39  & 69  & 39  & 37  & 30  & 29  & 27  & GT  & 60  & 0.00  & 1.69 \\
 4gt4-v0\_80  & 5  & 5  & 34  & 218  & 6  & 146  & 80  & 148  & 82  & 80  & 70  & 71  & 68  & GT  & 53  & 0.02  & 2.00 \\
 4gt5\_75  & 5  & 5  & 21  & 76  & 5  & 62  & 46  & 66  & 52  & 54  & 42  & 41  & 38  & GT  & 38  & 0.02  & 1.81 \\
 4mod5-v1\_23  & 5  & 8  & 24  & 90  & 5  & 74  & 44  & 75  & 57  & 54  & 42  & 42  & 40  & GLT  & 45  & 0.02  & 1.67 \\
 4mod7-v0\_95  & 5  & 6  & 38  & 144  & 5  & 112  & 72  & 112  & 72  & 144  & 69  & 67  & 66  & LT  & 41  & 0.02  & 1.74 \\
 add16\_174  & 49  & 64  & 192  & 220  & 49  & 382  & 254  & 413  & 413  & 382  & 496  & 496  & 254  & NT  & 33  & 0.54  & 1.32 \\
 add32\_183  & 97  & 128  & 384  & 444  & 97  & 766  & 510  & 829  & 829  & 766  & 1504  & 1504  & 510  & NT  & 33  & 2.32  & 1.33 \\
 add64\_184  & 193  & 256  & 768  & 892  & 193  & 1534  & 1022  & 1661  & 1661  & 1534  & 5056  & 5056  & 1022  & NT  & 33  & 14.07  & 1.33 \\
 add8\_172  & 25  & 32  & 96  & 108  & 25  & 190  & 126  & 205  & 205  & 190  & 184  & 184  & 126  & NT  & 33  & 0.11  & 1.31 \\
 aj-e11\_165  & 4  & 13  & 45  & 144  & 5  & 124  & 86  & 128  & 96  & 124  & 91  & 91  & 84  & GT  & 32  & 0.02  & 1.87 \\
 alu-v4\_36  & 5  & 7  & 31  & 136  & 5  & 102  & 70  & 106  & 76  & 94  & 59  & 56  & 54  & GT  & 47  & 0.03  & 1.74 \\
 cnt3-5\_180  & 16  & 20  & 120  & 1634  & 16  & 957  & 281  & 987  & 349  & 569  & 327  & 326  & 247  & GT  & 74  & 0.49  & 2.06 \\
 cycle10\_2\_110  & 12  & 19  & 1126  & 13472  & 12  & 7948  & 5372  & 7948  & 5372  & 7948  & 3490  & 3490  & 3430  & LT  & 56  & 3.61  & 3.05 \\
 decod24-v3\_46  & 4  & 9  & 9  & 36  & 4  & 27  & 15  & 27  & 15  & 19  & 13  & 14  & 13  & L  & 51  & 0.00  & 1.44 \\
 ham15\_108  & 15  & 70  & 453  & 9978  & 15  & 5470  & 3842  & 5458  & 3854  & 4982  & 1181  & 1168  & 1168  & GL  & 78  & 2.22  & 2.58 \\
 ham7\_104  & 7  & 23  & 83  & 624  & 7  & 403  & 285  & 411  & 299  & 277  & 172  & 169  & 167  & GT  & 58  & 0.09  & 2.01 \\
 hwb4\_52  & 4  & 11  & 23  & 40  & 4  & 51  & 41  & 59  & 51  & 51  & 37  & 37  & 37  & L  & 27  & 0.02  & 1.61 \\
 hwb5\_55  & 5  & 24  & 104  & 470  & 5  & 347  & 207  & 353  & 219  & 271  & 185  & 186  & 183  & LT  & 47  & 0.10  & 1.76 \\
 hwb6\_58  & 6  & 42  & 142  & 710  & 6  & 532  & 328  & 512  & 348  & 520  & 302  & 279  & 278  & GLT  & 47  & 0.16  & 1.96 \\
 hwb7\_62  & 7  & 331  & 2325  & 16890  & 8  & 11023  & 6883  & 11033  & 6921  & 10347  & 6164  & 6019  & 5985  & LT  & 45  & 3.90  & 2.57 \\
 hwb8\_118  & 8  & 633  & 14260  & 115030  & 9  & 72060  & 41938  & 72032  & 42014  & 70368  & 38801  & 38802  & 38801  & L  & 46  & 39.05  & 2.72 \\
 hwb9\_123  & 9  & 1959  & 18124  & 189426  & 10  & 115167  & 69663  & 115180  & 69720  & 114441  & 54970  & 54961  & 54961  & GL  & 52  & 73.77  & 3.03 \\
 mod5adder\_128  & 6  & 15  & 83  & 600  & 6  & 395  & 203  & 394  & 212  & 283  & 168  & 169  & 168  & L  & 57  & 0.08  & 2.02 \\
 mod8-10\_177  & 5  & 14  & 88  & 582  & 6  & 387  & 195  & 393  & 205  & 269  & 186  & 183  & 165  & GT  & 57  & 0.09  & 1.88 \\
 plus127mod8192\_162  & 13  & 910  & 57400  & 661596  & 14  & 396286  & 268836  & 396272  & 268866  & 396286  & 211476  & 211476  & 209194  & LT  & 47  & 269.17  & 3.64 \\
 plus63mod4096\_163  & 12  & 429  & 25492  & 254864  & 13  & 152998  & 102612  & 152984  & 102642  & 152998  & 87156  & 87156  & 87048  & LT  & 43  & 60.56  & 3.41 \\
 plus63mod8192\_164  & 13  & 492  & 32578  & 397864  & 14  & 236066  & 161188  & 236052  & 161214  & 236066  & 117740  & 117740  & 115070  & LT  & 51  & 132.34  & 3.53 \\
 rd32-v0\_67  & 4  & 2  & 10  & 10  & 4  & 18  & 14  & 19  & 17  & 12  & 16  & 12  & 12  & G  & 33  & 0.02  & 1.20 \\
 rd53\_135  & 7  & 16  & 77  & 466  & 7  & 326  & 202  & 314  & 216  & 286  & 162  & 153  & 153  & GL  & 53  & 0.11  & 1.99 \\
 rd73\_140  & 10  & 20  & 76  & 450  & 10  & 314  & 176  & 315  & 197  & 266  & 152  & 149  & 138  & LT  & 56  & 0.08  & 1.82 \\
 rd84\_142  & 15  & 28  & 112  & 910  & 15  & 580  & 274  & 581  & 299  & 640  & 260  & 270  & 260  & L  & 55  & 0.16  & 2.32 \\
 sym9\_148  & 10  & 210  & 4368  & 48736  & 10  & 28820  & 18338  & 28820  & 18338  & 25444  & 9849  & 11309  & 9848  & LT  & 65  & 8.88  & 2.25 \\
 sys6-v0\_144  & 10  & 15  & 67  & 358  & 10  & 254  & 150  & 255  & 169  & 322  & 129  & 144  & 129  & L  & 49  & 0.03  & 1.93 \\
 urf1\_149  & 9  & 11554  & 57770  & 462708  & 9  & 303374  & 156424  & 300962  & 165226  & 258230  & 118005  & 118010  & 118005  & L  & 61  & 193.37  & 2.04 \\
 urf2\_152  & 8  & 5030  & 25150  & 171284  & 8  & 115826  & 61302  & 115666  & 65100  & 115826  & 50661  & 50661  & 50652  & LT  & 56  & 35.66  & 2.01 \\
 urf3\_155  & 10  & 26468  & 132340  & 1282724  & 10  & 795496  & 387346  & 799448  & 410372  & 732684  & 287016  & 287017  & 287012  & LT  & 63  & 831.05  & 2.17 \\
 urf5\_158  & 9  & 10276  & 51380  & 442748  & 9  & 280948  & 145478  & 280052  & 151332  & 256748  & 103823  & 103822  & 103820  & LT  & 63  & 99.52  & 2.02 \\
 urf6\_160  & 15  & 10740  & 53700  & 951276  & 15  & 531768  & 232354  & 531664  & 234208  & 480772  & 142604  & 142603  & 142600  & LT  & 73  & 302.61  & 2.66 \\
\hline
\end{tabular}
\textsc{N}: Naive method (i.e., synthesis, decomposition, SWAP insertion)\hspace{.3cm}
\textsc{T}: With template matching \hspace{.3cm}
\textsc{M}: With macros replacement \hspace{.3cm}
\textsc{G}: With global reordering\hspace{.3cm}
\textsc{L}: With local reordering \hspace{.3cm}

\begin{flushleft}
Column \emph{Time} denotes the overall run-time needed to generate the results for \emph{all} possible configurations.\\
Almost all results have been obtained in negligible run-time (i.e., in less than one CPU second). Only if template matching was enabled more run-time was needed.

\end{flushleft}

\end{sidewaystable}

\begin{figure}[t]
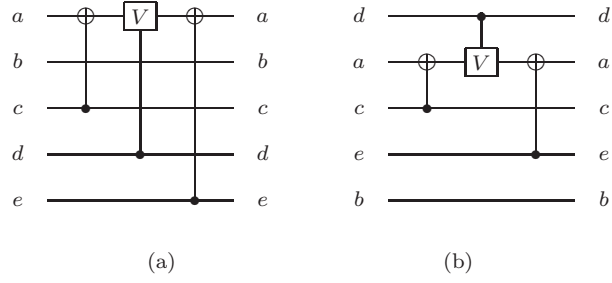

        \centering
        \subfigure[\label{fig:motivation_reordering1}]{
          \input{qcircuits/reordering_motivation1.tex}
         }\subfigure[\label{fig:motivation_reordering2}]{
          \input{qcircuits/reordering_motivation2.tex}
        }
        \caption{Reordering circuit lines}
\end{figure}

Applying the approaches introduced so far leads to an increase in the quantum cost for each non-adjacent gate. In contrast, by modifying the ordering of the circuit lines, some of the additional costs can be saved. As an example, consider the circuit in Fig.~\ref {fig:motivation_reordering1} with quantum cost 3 and an NNC value of 6. By reordering the lines as shown in Fig.~\ref{fig:motivation_reordering2}, the NNC value can be reduced to 1 without increasing the total quantum cost.
It is worth noting that manipulating the line order has been previously done to reduce the quantum cost (e.g.,~in~\cite{MDM:2005,WGDD:2009}).
To determine which lines should be reordered, two heuristic methods are proposed in the following. The former one changes the ordering of the primary inputs and outputs according to a global view while the latter one applies a local view to assign the line ordering.

\subsubsection{Global Reordering}
After applying the standard decomposition algorithms \cite{Barenco95,MaslovTCAD08}, a cascade of 1- and 2-qubit gates is generated. Now, an ordering of the circuit lines which reduces the total NNC value is desired. To do that, the ``contribution'' of each line to the total NNC value is calculated. More precisely, for each gate $g$ with control line $i$ and target line $j$, the NNC value is calculated. This value is added to variables $imp_i$ and $imp_j$ which are used to save the impacts of the circuit lines $i$ and $j$ on the total NNC value, respectively. Next, the line with the highest NNC impact is chosen for reordering and placed at the middle line (i.e.,~swapped with the middle line). If the selected line is the middle line itself, a line with the next highest impact is selected. This procedure is repeated until no better NNC value is achieved. Finally, SWAP operations as described in the previous sections are added for each non-adjacent gate. The following example illustrates the idea.

\begin{figure}[t]
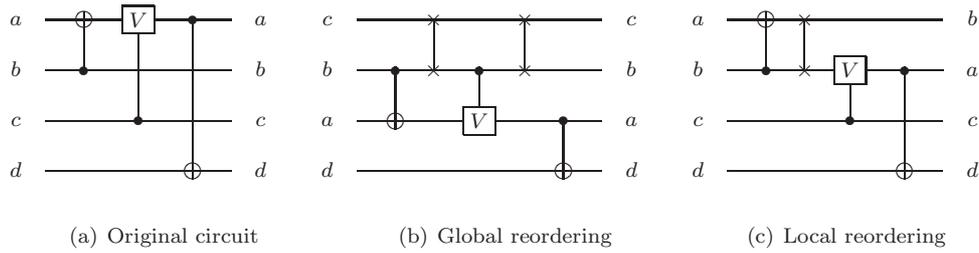

        \centering
        \subfigure[Original circuit\label{fig:reordering1}]{
          \input{qcircuits/reordering_local1.tex}
        }\subfigure[Global reordering\label{fig:reordering_global2}]{
          \input{qcircuits/reordering_global2.tex}
        }\subfigure[Local reordering\label{fig:reordering_local2}]{
          \input{qcircuits/reordering_local2.tex}
        }
        \caption{Global and local reordering}
\end{figure}

\begin{example}\label{example:global_reordering}
Consider the circuit depicted in Fig.~\ref{fig:reordering1}. After calculating the NNC contributions, we have $imp_a=1.5$, $imp_b=0$, $imp_c=0.5$, and $imp_d=1$, respectively. Thus, lines $a$ (highest impact) and $c$ (middle line) are swapped. Since further swapping does not improve the NNC value, reordering terminates and SWAP gates are added for the remaining non-adjacent gates. The resulting circuit is depicted in Fig.~\ref {fig:reordering_global2} and has quantum cost of 9 in comparison to 21 that results if the naive method is applied.
\end{example}

\subsubsection{Local Reordering}
In order to save SWAP gates, line ordering can also be applied according to a local schema as follows. The circuit is traversed from the inputs to the outputs. As soon as there is a gate~$g$ with an NNC value greater than~0, a SWAP operation is added in front of~$g$ to enable an adjacent gate. However, in contrast to the naive NNC-based decomposition, no SWAP operation is added after~$g$. Instead, the resulting ordering is used for the rest of the circuit (i.e.,~propagated through the remaining circuit). This process is repeated until all gates are traversed.

\begin{example} \label {exm:localordering}
Reconsider the circuit depicted in Fig.~\ref{fig:reordering1}. The first gate is not modified since it has an NNC of 0. For the second gate, a SWAP operation is applied to make it adjacent. Afterwards, the new line ordering is propagated to all remaining gates resulting in the circuit shown in Fig.~\ref{fig:reordering_local2}. This procedure is repeated until the whole circuit has been traversed. Finally, a circuit with quantum cost of 9 (in contrast to 21) is produced.
\end{example}

\section{A Synthesis Flow for LNN Architectures}\label{sec:integration}

\begin{figure}[h]
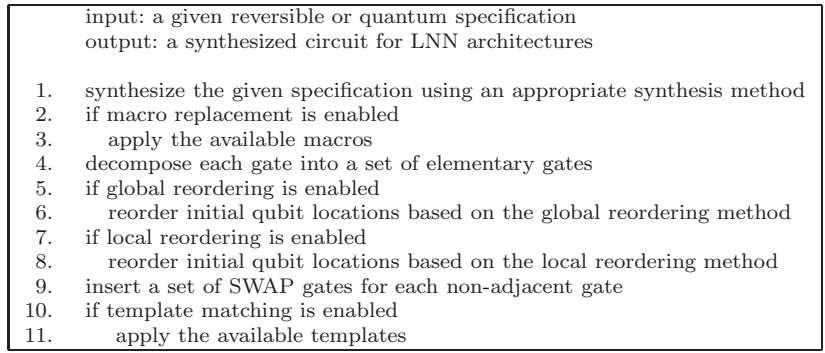

    \begin{center}
       \begin{tabular}{|rl|} \hline
       &input: a given reversible or quantum specification \\
       &output: a synthesized circuit for LNN architectures \\
       &\\
       1. &synthesize the given specification using an appropriate synthesis method \\
       2. &if macro replacement is enabled \\
       3. &~~~apply the available macros \\
       4. &decompose each gate into a set of elementary gates\\
       5. &if global reordering is enabled \\
       6. &~~~reorder initial qubit locations based on the global reordering method\\
       7. &if local reordering is enabled \\
       8. &~~~reorder initial qubit locations based on the local reordering method\\
       9. &insert a set of SWAP gates for each non-adjacent gate \\
       10. &if template matching is enabled \\
       11. &~~~ apply the available templates\\
       \hline
       \end{tabular}
    \end{center}
    \caption{The extended synthesis flow}
    \label{flowdetails}
\end{figure}

Having the proposed approaches from the previous section available, they can be combined to an extended synthesis flow that explicitly takes the LNN limitation into account. Fig.~\ref{flowdetails} illustrates this flow. As shown in this figure, first an off-the-shelf synthesis approach is applied to create an initial circuit realization. Afterwards, if macro replacement is enabled, the proposed macro replacement method from Section~\ref{sec:opt_exact} is applied to simplify the circuit (lines 2-3). Then, one of the available standard decomposition methods is applied to decompose all non-elementary gates into a set of elementary unit-cost gates (line 4). The resulting quantum circuit can be optimized by the reordering methods proposed in Section~\ref{sec:opt_reordering} (lines 5-8). Finally, SWAP gates for the remaining non-adjacent gates have to be added (line 9) and template matching as introduced in Section~\ref{sec:TM} can additionally be applied (lines 10-11). Note that each method is applied on the result of the previous method in the proposed synthesis flow. It can be verified that for the naive method, only lines 1, 4, and 9 are executed.

\section {Experimental Results} \label {sec:exp}

In this section, experimental results are presented. We evaluated the methods introduced in Section \ref{sec:opt} and compared them to the naive approach, which has been used by other synthesis methods~\cite{Hirata09,mottonen06,Chakrabarti07,Khan08,Shende06,SaeediJETC10} so far. All approaches have been implemented in C++ and applied to the benchmark collection available at RevLib \cite{WilleMVL08} including a wide variety of circuits that already have been used by other researchers to evaluate previous reversible synthesis approaches. The experiments have been carried out on an Intel Pentium IV 2.2GHz computer with 2GB memory.

The results are shown in Table \ref{table:nnc} and Table~\ref{table:nnc2}, respectively.
The former table shows the results obtained by applying the established decomposition, where a SWAP gate is composed of three elementary gates. Additionally, Table~\ref{table:nnc2} shows the results obtained by assuming the SWAP gate itself to be an elementary gate (as done by certain quantum technologies~\cite{Nielsen00}).
The first column gives thereby the names of the circuits followed by unique identifiers as used in RevLib. Then, the number of circuit lines~(\emph{n}), the gate count~(\emph{gc}), the quantum cost~(\emph{qc}), and the NNC value of the original reversible circuits are shown. The following columns denote the quantum cost of the NNC-optimal circuits obtained by the naive method (N) as well as by the proposed synthesis flow where a combination of macro replacement (\emph{M}), global reordering (\emph{G}), local reordering (\emph{L}), and template matching (\emph{T}) methods has been applied. For example, \emph{MT} denotes the results obtained by the proposed flow with macro replacement and template matching methods enabled (i.e., reordering methods disabled). Besides that, the results of the best configurations are given in column~\emph{Best config}. The percentages of the best quantum cost reduction obtained by the extended synthesis flow in comparison to the widely used naive method are reported in Column~\emph{Best Impr.}. Column \emph{Time} denotes the overall run-time needed to generate the results for \emph{all} possible configurations (with and without any possible options). Finally, the last column shows the remaining overhead in terms of quantum cost needed to achieve NNC-optimality in comparison to the original circuit (\emph{Ohead}).

As can be seen, decomposing reversible circuits to have NNC-optimal quantum circuits for LNN architectures is costly. Using the widely used naive method, the quantum cost increases significantly. 
This result has been obtained in recent synthesis papersas well \cite{Cheung07,Shende06,SaeediJETC10}. However, using the proposed methods, this can be improved. Even if reordering may worsen the results in some few cases,
in total this leads to an improvement. The results have been obtained in negligible run-time (i.e.,~in less than one CPU second). Only if template matching was enabled more run-time was needed.

Overall, reductions of more than 50\% on average -- in the best case of 83\% -- have been observed considering the established decomposition (see Table \ref{table:nnc}). Similar results are obtained applying the extended definition of elementary gates (see Table \ref{table:nnc2}).
As a result, NNC-optimal circuits can be synthesized with a moderate increase of quantum cost.

\section{Conclusions} \label {sec:conclusion}
Quantum technologies are in preliminarily state and several limitations should be resolved to have a scalable quantum technology. Limited interaction distance between gate qubits is one of the most common limitations of the current technologies. In this paper, we illustrated how the synthesis flow can be modified to produce efficient circuits for quantum technologies with limited interactions. The proposed flow includes a set of NNC-based decomposition methods equipped by an NNC-based template matching algorithm. The experiments show that with a naive treatment of the LNN restriction, quantum circuits require up to one order of magnitude higher quantum cost in the LNN architectures. In contrast, using the suggested methods, this increase can be reduced by more than 50\% on average (83\% in the best case).

\section*{Acknowledgment}
Parts of this research work has been supported by the German Research Foundation (DFG) (DR 287/20-1). The final publication is available at www.springerlink.com:\\
\texttt{http://www.springerlink.com/content/v250758vh5053qxx/}

\end{document}